\documentclass[12pt,a4paper]{article}
\usepackage{dfttrob}  
\usepackage{amsmath}
\usepackage{graphicx}

\dfttnum{DFTT 09/2003\\April 2003}  

\begin{document}

\title{On signals of new physics in global event properties in pp 
collisions in  the TeV energy domain}
\author{A. Giovannini and R. Ugoccioni\\
 \it Dipartimento di Fisica Teorica and INFN - Sezione di Torino\\
 \it Via P. Giuria 1, 10125 Torino, Italy}
\maketitle

\begin{abstract}
In the framework of the weighted superposition mechanism
of different classes of minimum bias events (or substructures),
described by the negative binomial multiplicity distribution,
in possible scenarios for pp collisions in the TeV energy domain,
we explore global properties of an eventual new class of
events, characterised by high hadron and clan densities, to be added to
the soft (without minijets) and semihard (with minijets) ones.
It turns out that the main signal of the mentioned new physical
expectations at 14 TeV c.m.\ energy would be an
``elbow structure'' in the tail of the total charged particle multiplicity
distribution in complete disagreement with the second shoulder structure
predicted by Pythia Monte Carlo calculations: a challenging problem
for new experimental work.
\end{abstract}

\newcommand{\nbar}{\bar n}            
\newcommand{\Nbar}{\bar N}            
\newcommand{\nc}{{\bar n_c}}          

\newcommand{\III}{\ensuremath{\text{th}}}
\newcommand{\nbarsoft}{\nbar_{\text{soft}}}
\newcommand{\nbarsemi}{\nbar_{\text{sh}}}
\newcommand{\nbariii}{\nbar_{\III}}
\newcommand{\ksoft}{k_{\text{soft}}}
\newcommand{\ksemi}{k_{\text{sh}}}
\newcommand{\kiii}{k_{\III}}
\newcommand{\Nbarsoft}{\Nbar_{\text{soft}}}
\newcommand{\Nbarsemi}{\Nbar_{\text{sh}}}
\newcommand{\Nbariii}{\Nbar_{\III}}
\newcommand{\ncsoft}{\nbar_{c,\text{soft}}}
\newcommand{\ncsemi}{\nbar_{c,\text{sh}}}
\newcommand{\nciii}{\nbar_{c,\III}}
\newcommand{\aiii}{a_{\III}}
\newcommand{\biii}{b_{\III}}
\newcommand{\piii}{p_{\III}}
\newcommand{\alphasoft}{\alpha_{\text{soft}}}
\newcommand{\alphasemi}{\alpha_{\text{sh}}}
\newcommand{\alphaiii}{\alpha_{\text{th}}}

\section{Introduction}

Possible scenarios for pp collisions in the TeV energy domain, based on
extrapolations from data and fits in the GeV region to
charged particle multiplicity distributions, $P_n$, and on the
weighted superposition mechanism of two classes of events, i.e., soft
(without minijets) and semihard (with minijets), each described by a
negative binomial multiplicity distribution (NBMD) with characteristic
parameters ($\nbarsoft, \ksoft$ and $\nbarsemi, \ksemi$, respectively)
have been discussed both in full phase space \cite{combo:prd} and rapidity
intervals \cite{combo:eta}. (For an experimental investigation of the
substructures see Ref.~\cite{CDF:soft-hard}.)

It has been pointed out that in the semihard component either with a
strongly KNO scaling violating mechanism (parameter $\ksemi$ decreases
with c.m.\ energy $\sqrt{s}$ as $\sim 1/\log\sqrt{s}$) or with a QCD
inspired KNO scaling violating behaviour, (which is milder than the
previous one as parameter $\ksemi$ decreases as $\sim 1/\sqrt{\log
s}$), the average number of clans, $\Nbarsemi$, becomes smaller
as the c.m.\ energy increases (from $\approx 23$ and 22, resp., at 900 GeV, 
to $\approx 11$ and 18, resp., at 14 TeV)
and the corresponding average number of particles per clan, $\ncsemi$,
much larger (from $\approx 2.5$ and 2.6, resp., at 900 GeV, 
to $\approx 7$ and 5, resp., at 14 TeV), favouring clan
aggregation and higher average particle population within clans. This
behaviour was considered suggestive and indicative of  a
class of events with high particle density.

Motivated by this remark, we decided to explore in the same framework
the consequences of an eventual more dramatic decrease of the average
number of clans towards 1, i.e., towards a class of events with the
maximum clan and particle aggregation, a situation which implies a  
$\ksemi$ value less than one but which is unfortunately met in the
semihard component only asymptotically at an extraordinary high c.m.\
energy in the KNO scaling violating extrapolations of $\ksemi$.
This result contrasts sharply with the main properties of the class of
events belonging to the semihard component at non-asymptotic energies,
where $\ksemi$ is always larger than or equal to one.  It raises the
intriguing question whether a very high clan
and particle density is only an asymptotic behaviour of the semihard
component, or the property of an effective new class of events, 
different from (maybe harder than) the semihard one, 
and whose onset might happen
already at 14 TeV c.m.\ energy: if this is the case, its contribution
to the total charged particle multiplicity distribution,
$P_n^{(\text{total})}$, in terms of a new weighted NB(Pascal)MD, should
be added to those of the soft and semihard ones of Ref. \cite{combo:prd} and
\cite{combo:eta}.
The claim is that the benchmark of this third new class of events (called 
`\III'\ from now on)  is  $\kiii < 1$.

\begin{figure}
  \begin{center}
  \mbox{\includegraphics[width=0.7\textwidth]{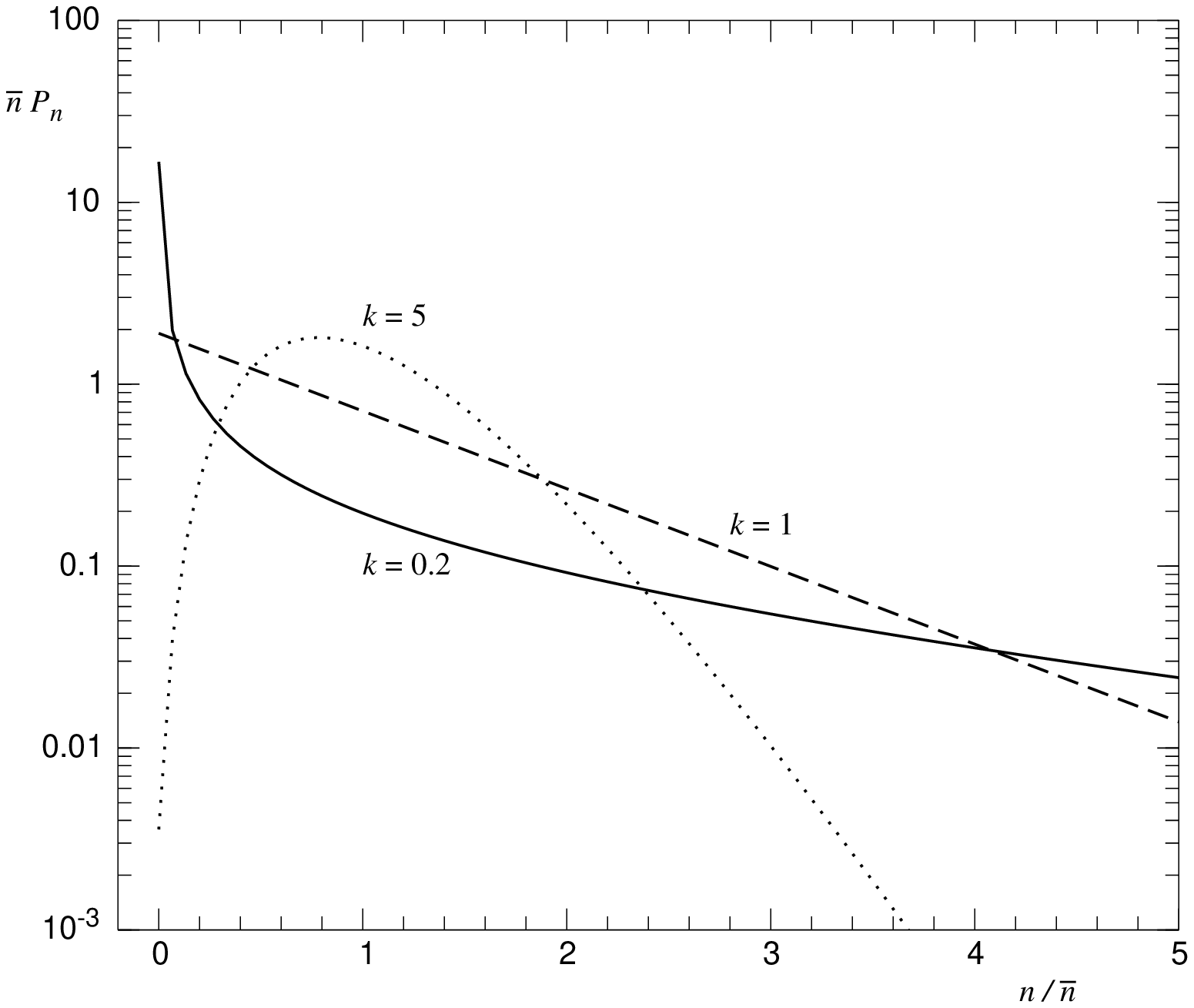}}
  \end{center}
  \caption{Multiplicity distributions, in KNO form, with various 
		values of $k$ and the same average multiplicity $\nbar$, show
		different curvatures.}\label{fig:gamma}
  \end{figure}

It should be recalled that
the average number of clans  is in general  a non trivial function 
of the average number of particles of each  component, $\nbar_i$,  and 
of the parameter $k_i$,
\begin{equation}
	       \Nbar_i =  k_i \ln ( 1 + \nbar_i / k_i )        \label{eq:1}
\end{equation}
with $i$=soft, sh and \III, enumerating the three classes.
Whereas $\ksoft$ is taken constant in our scenarios 
in the new energy domain and therefore $\Nbarsoft$ depends only on 
the average multiplicity $\nbarsoft$
as the c.m.\ energy increases, the decrease of $\ksemi$ with c.m.\ energy  
suggested by our assumptions on  KNO scaling violations  remains at   
non-asymptotic energies  always  larger than  or equal to one in the 
semihard  component and  is  only in part contrasted by the increase of 
$\nbarsemi$, a fact which  outlines  $\ksemi$ dominance in the general  
$\Nbarsemi$ behaviour. As already mentioned the request of an average 
number of clans, $\Nbariii$, of few units in the third component  leads  to 
values of  $\kiii$ less than one  whereas  that of  
$\Nbariii \approx 1$  implies $\kiii \to 0$ for large $\nbariii$.  
Notice that  in this last extreme 
situation the average number of particle per clan, $\nciii$ coincides 
of course with the  average number of particles of the new component $\nbariii$.
In addition when  $\nbarsemi \gg \ksemi$ ---a quite normal situation
at large c.m.\ energies---  the NB(Pascal)MD 
describing the final charged particle 
MD of the component is well approximated by a log-concave gamma MD for 
$\ksemi$ larger than one which becomes  an exponential when $\ksemi$ is
equal to one:
the maximum of the gamma distribution occurs at  $n/\nbariii = 1-\kiii^{-1}$.
(See Fig.~\ref{fig:gamma}).
This behaviour should be compared with what happens in the third component
under the condition $\kiii \ll \nbariii$:
in this case the NBMD is in fact  well approximated by a log-convex gamma
distribution  which for a certain value of the above parameter close to
zero leads to the average number of clans $\approx 1$ and to the total MD
well described by a logarithmic one. This  result is consistent  
with the standard   interpretation of the occurrence of the negative binomial 
 MD for final charged multiplicity at hadron level  as a two step process 
in which the  independently
emitted  clans contain logarithmically distributed charged  particles
\cite{AGLVH:4}. 

Finally  in  consequence of the assumed quite extreme clan and particle 
aggregation into few clans,  events with quite large   
forward-backward multiplicity   correlations close to the maximum
allowed leakage from one hemisphere to the opposite one 
\cite{RU:FB} are also expected. 
At partonic level the mentioned remarks  would suggest for the third 
component  high parton density clan production with huge colour exchange 
processes originated from a relatively 
small number of  high virtuality ancestors, which would indicate probably 
an emission mechanism harder  than that seen in the semihard component.

In conclusion  it might well be that one could observe 
already at 14 TeV c.m.\ energy in pp collisions  three 
classes of events or components instead of two,  the first and the second 
class being  those  examined in Refs \cite{combo:prd} and \cite{combo:eta} 
and the third  one  fully 
characterised by the reduction of the the average number of 
clans to the minimum 
allowed by the condition  $\kiii < 1$. The classes of events or
components contributing to the total $n$ charged particle multiplicity
distributions would therefore be the following:  

\noindent \textit{I}) the  class of soft events (events  with no minijets), 
which in clan structure
analysis would be characterised by quite large values of  $\Nbarsoft$ and 
quite small $\ncsoft$; $P_n^{(\text{soft})}$ obeys KNO scaling 
and $\ksoft$ is assumed to
be constant from the GeV region to  the new energy domain;

\noindent \textit{II}) the class of semihard events (events with minijets),
for which $\Nbarsemi$ decreases quickly with c.m.\ energy  and 
$\ncsemi > \ncsoft$; $\ksemi$ decreases also  with c.m.\ energy
but its value is larger than or equal to one   and   KNO scaling is violated;

\noindent \textit{III}) the new  class of
events (events generated probably by quite hard partons) with
$\kiii < 1$, $\nbariii \gg \kiii$ (and quite small $\Nbariii$), with large 
forward-backward multiplicity correlations.

Accordingly, the total charged particle MD will have  the following
expression in terms of the weighted composition of the  three NBMD's,  one for
each component:
\begin{equation}
	P_n^{(\text{total})} = \alphasoft P_n^{(\text{soft})} + 
	       \alphasemi P_n^{(\text{sh})} +
       (1 - \alphasoft - \alphasemi ) P_n^{(\III)}         \label{eq:2}
\end{equation}
where $\alphasoft$ and $\alphasemi$ are the fraction of soft and
semihard events, respectively.

At 14 TeV c.m energy  the weight of the third component is expected
to be of course  quite small, in our calculations we assumed it to be
between 1 and 3 percent of the total number of events.

Since the quantitative  properties of the classes of soft and semihard events 
have been extensively discussed in Refs. \cite{combo:prd} and \cite{combo:eta} 
we focus our attention 
in this paper on the global  properties  of the new  class of  events, which
we would be tempted to call hard events.

\section{A new  class of events, a third component in pp collisions?}

Assuming for simplicity that the new class of events is described
---as are the soft and semihard ones--- by a NBMD with parameters
$\nbariii$ and $\kiii$, we decided to examine under such conditions
possible signals of new physics which should be easily detectable
already at 14 TeV c.m.\ energy in pp collisions by studying the
general behaviour of the various sets of parameters characterising the
final charged particles MD of the component $P_n (\nbariii, \kiii)$.
In order to stress the properties of the new class of events, the
extreme case $\Nbariii = 1$ will be discussed in the following.

\subsection{Consequences in terms of the $\nbariii$, $\kiii$ parametrisation of the NBMD.}

\begin{figure}
  \begin{center}
  \mbox{\includegraphics[width=0.7\textwidth]{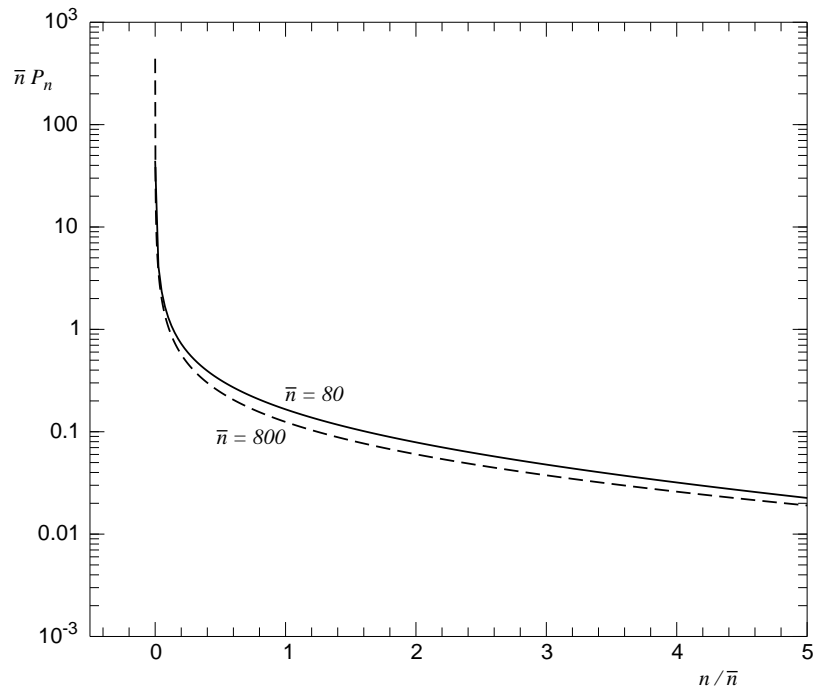}}
  \end{center}
  \caption{Multiplicity distributions in KNO form for two values of
		$\nbar$, with the respective values of $k$ (0.1611 for $\nbar=80$
		and 0.1128 for $\nbar=800$)
    obtained from Eq.~(\ref{eq:3}), i.e., requiring $\Nbar=1$.}\label{fig:N1}
  \end{figure}

The request that $\Nbariii = 1$, through its definition 
Eq.~(\ref{eq:1}),  leads to the straightforward relation between
$\nbariii$ and $\kiii$:
\begin{equation}
   \nbariii = \kiii ( e^{1/\kiii}  - 1)       \label{eq:3}	
\end{equation}
with (being $\nbariii \gg \kiii$)
\begin{equation}
	\kiii < 1.                         \label{eq:4}
\end{equation}
In Fig.~\ref{fig:N1} are plotted $\nbariii P_n$ vs $n/\nbariii$ in KNO
form for $\Nbariii = 1$ and different average multiplicities, with the
respective values of $\kiii$ obtained via Eq.~(\ref{eq:3}),
which highlights the general trend of the exponential behaviour of the
MD and $\kiii < 1$ dominance.
In addition we notice that for $\kiii < 1$, as for the gamma
distribution corresponding to the limit $\nbar \gg k$ of the NBMD,
we can distinguish two regions: $n > \nbariii$ and $n < \nbariii$.
In the former a smooth behaviour of $\nbariii P_n$ is seen as $n$
increases (events with large multiplicities) whereas in the
latter a dramatic decrease of $\nbariii P_n$ is visible (events with
low multiplicities).
The fact that the distribution decreases so slowly for $n > \nbariii$ 
and so fast for $n < \nbariii$ produces more easily
events with very large or very small charged multiplicity, which
reminds us of what occurs in cosmic rays for centauro and
anti-centauro events.
A provocative result to be tested in experiments.

\subsection{Consequences in terms of clan structure analysis}\label{sec:N+nc}

\begin{figure}
  \begin{center}
  \mbox{\includegraphics[width=0.7\textwidth]{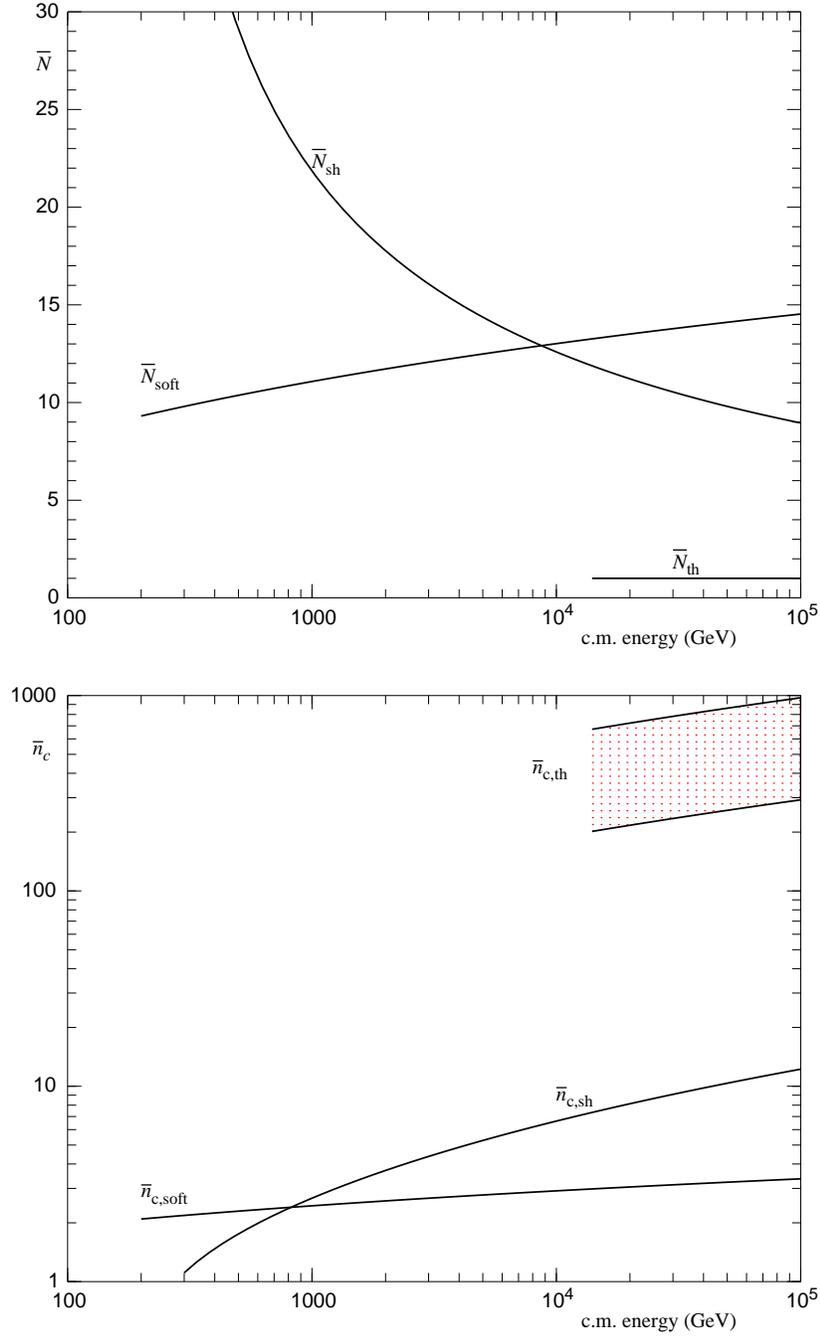}}
  \end{center}
  \caption{Energy dependence of the average number of clans (top panel) and 
	 of the average number of particles per clan (bottom panel) for the
	 three components (the band illustrates different choices for
	 $\nbariii$, see discussion in section \ref{sec:N+nc}).}\label{fig:Nclan}
  \end{figure}

In Fig.~\ref{fig:Nclan} are plotted $\Nbar$ and $\nc$ for the three
components as a function of c.m.\ energy in full phase space:
the first two components are those discussed in \cite{combo:prd},
notice that $\ksemi$ behaviour corresponds to that suggested by a
strong KNO scaling violation.
The lack of experimental data on the third component at lower c.m.\
energies and the consequent impossibility to guess its behaviour in terms of
extrapolations as it was done for the soft and semihard components in
Ref.~\cite{combo:prd}, led us to show a band of possible values.
A conservative guess that the total multiplicity variation due to the
third component over the extrapolation of Ref.~\cite{combo:prd} is
limited to 10\% leads to $\nbariii$ values from 3 to 10 times larger
than the total multiplicity, i.e., to a third component weight ranging
from 1 to 3\%.
Notice that to a small variation of $\kiii$ according to
Eq.~(\ref{eq:3}) corresponds a quite large variation of $\nbariii$.
This is clearly shown in the figure, where $\Nbariii = 1$ whereas
$\nciii$ is 3 times larger in one case with respect to the other.
Last but not least, $\nciii$ is, as assumed, much larger than in the
two other components.

\subsection{Consequences in terms of parameters $\aiii, \biii$ of the NBMD.}\label{sec:a+b}

Since 
\begin{equation}
	\frac{ (n+1) P_{n+1}^{(\III)} }{ P_{n}^{(\III)} } =
	          \aiii + \biii n ,
\end{equation}
with
\begin{equation}
   \aiii = \frac{ \kiii  \nbariii }{  \kiii + \nbariii  }, 
\end{equation}
and
\begin{equation}
   \biii = \frac{ \nbariii }{  \kiii + \nbariii  },
\end{equation}
it follows that
\begin{equation}
	\aiii = \Nbariii P_{\text{log}}(1)
           =  \Nbariii  \frac{- \biii }{ \ln(1-\biii) },
\end{equation}
(where $P_{\text{log}}$ is the logarithmic distribution)
which for $\Nbariii = 1$ leads to 
\begin{equation}
  \aiii = - \biii / \ln (1-\biii) = P_{\text{log}}(1) .
\end{equation}
It should be noticed that in  the limit    $\kiii \to 0$
(it corresponds to  $\aiii \to 0$   and $\biii \to 1$) a   
truncated NBMD (i.e., with zero multiplicity missing)  and with  constant 
$\nbariii / \kiii$ (i.e., constant  $\biii$)
leads to a logarithmic MD, i.e., to  the MD of a single average clan:
\begin{equation}
  P_n^{(\III)} \to \frac{(\biii)^{n-1} }{ n }  P_{\text{log}}(1)
          \stackrel{\aiii \to 0}{=}
             -\frac{(\biii)^n }{ n  \ln (1-\biii) }
					= P_{\text{log}}(n)  .
\end{equation}
This result is of course consistent with the standard interpretation
of the occurrence of the NBMD as a two step process (independently
produced clans with at least one particle ancestor decay according to
a logarithmic MD) when the average number of clans is reduced to one.

\begin{figure}
  \begin{center}
  \mbox{\includegraphics[width=0.7\textwidth]{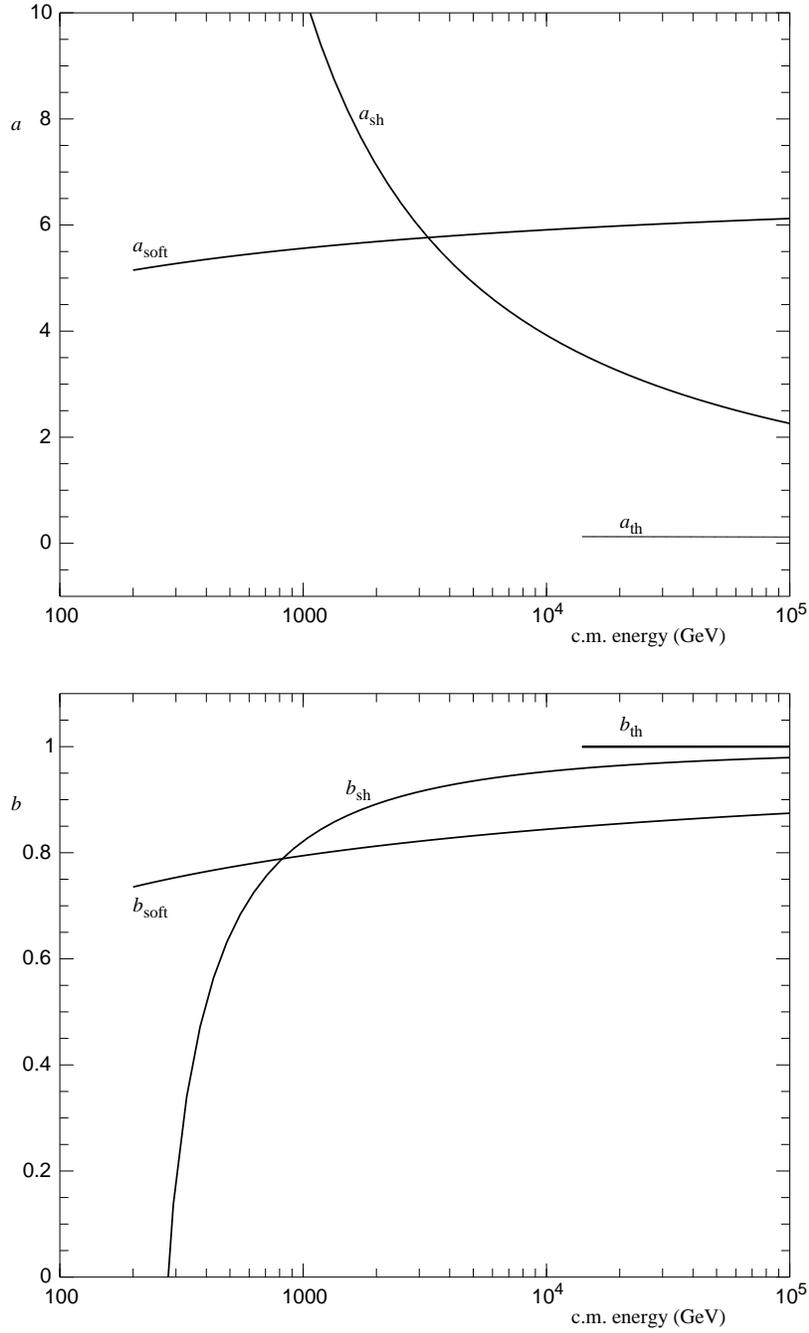}}
  \end{center}
  \caption{Energy dependence of $a$ (top panel) and $b$ (bottom panel)
  parameters for the three components (see discussion in section
	\ref{sec:a+b}; no band is visible for these variables, as they are
  little sensitive to the exact value of $\nbariii$).}\label{fig:ab}
  \end{figure}

In Fig.~\ref{fig:ab} are plotted the c.m.\ energy dependence of $a$
and $b$ for the three components which highlight the general trend
of $\aiii \to 0$ and $\biii\to 1$ for $\Nbariii\to 1$.
Notice that in general, for $\nbar_i \gg k_i$, one has $a_i \sim k_i$.

\subsection{Clan  aggregation and correlations.}

Being the probability of
two particles of the hard component to join the same clan
$1/\kiii$ times larger than the probability to join two different
clans \cite{AGLVH:4} it is clear that $\kiii < 1$ is 
enhancing particle aggregation
properties within clans.

Being $\nbariii^2 / \kiii$   much larger than in the semihard
component,  from the relation
\begin{equation}
	\nbariii^2 / \kiii  = \int C_2 (\eta_1',\eta_2'') d \eta_1' d \eta_2'' 
\end{equation}
two particle correlations are also expected to be larger.
Since cumulants are dependent on $1/\kiii$, which is again much  larger
than $1 / \ksemi$,  higher order cumulants are also expected to be
quite large in the third component.

Concerning forward-backward multiplicity correlations and 
leakage parameter behaviour, it should be pointed out,
as discussed in Ref.~\cite{RU:FB}, that the forward-backward multiplicity
correlation strength is given for each component by
\begin{equation}
   b_{\text{FB},i} =  \frac{ 2 b_i  p_i ( 1- p_i  ) }{ 1  - 
	 2  b_i  p_i (1-p_i) },
\end{equation}
where $p_i$ is the average fraction of particles, within each clan,
which remain in the same hemisphere where the clan was emitted and do
not `leak' to the opposite hemisphere.
In the third component, in presence of only one clan, 
the leakage is expected to be
maximum ($\piii \to 1/2$); being at the same time, 
as we have seen, $\biii \to 1$, one obtains $b_{\text{FB},\III}\to 1$.

Accordingly forward-backward multiplicity correlations in the third
component are expected to be quite stronger than in the semihard component.

\subsection{What Monte Carlo event generators say on the third component at
14 TeV c.m.\ energy in pp collisions}

\begin{figure}
  \begin{center}
  \mbox{\includegraphics[width=0.7\textwidth]{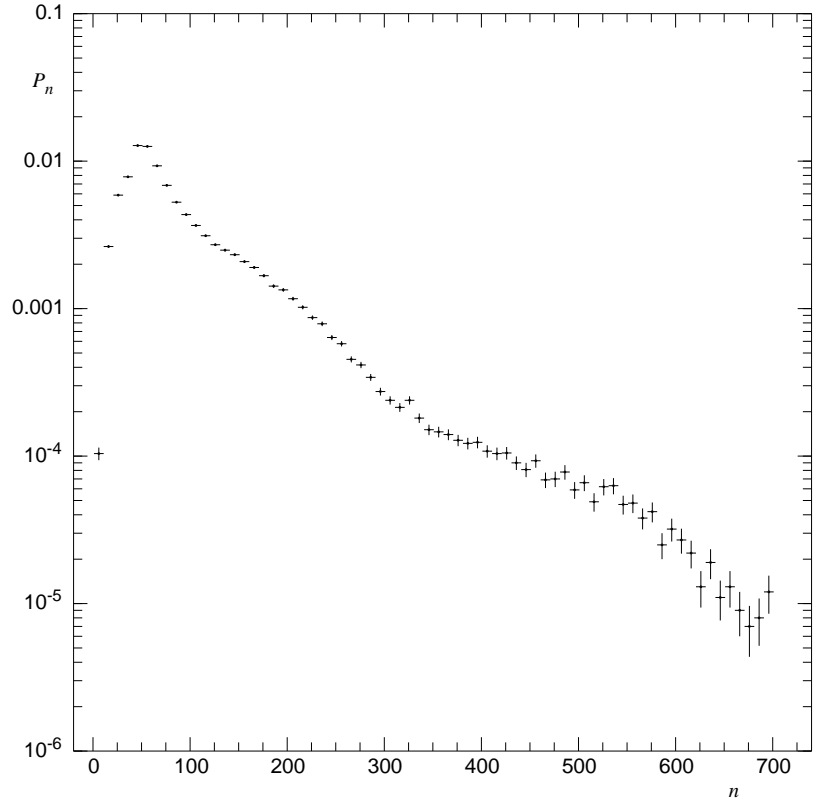}}
  \end{center}
  \caption{$n$ charged particle multiplicity distribution $P_n$
    predicted for minimum bias events
		in full phase space by 
		Pythia Monte Carlo (version 6.210, default parameters using
    model 4 with a double Gaussian matter distribution)
		at 14 TeV c.m.\ energy, showing two shoulder structures.}\label{fig:mdpy}
  \end{figure}

It is interesting to point out that the total charged particle  MD, $P_n$,
of the events generated with the last version of the Pythia Monte
Carlo model \cite{Pythia}
both in full phase space
and in a restricted rapidity interval ($|\eta|< 0.9$)
cannot be fitted in terms of the weighted superposition of two NBMD's
and that  the plot of $P_n$ vs $n$ shows two shoulders (see
Fig.~\ref{fig:mdpy}). 
These two shoulders appearing in the total multiplicity  distribution
come, in our framework, from
the weighted  superpositions of the first with the second  and of the second
with the third component. 
This is true also in the pseudo-rapidity interval.  
It would be tempting to identify the third component with the
hard one discussed above, but striking differences between the 
third class of events
generated with  Pythia Monte Carlo and that one  discussed above prevent us
from doing this identification.

\begin{figure}
  \begin{center}
  \mbox{\includegraphics[width=0.7\textwidth]{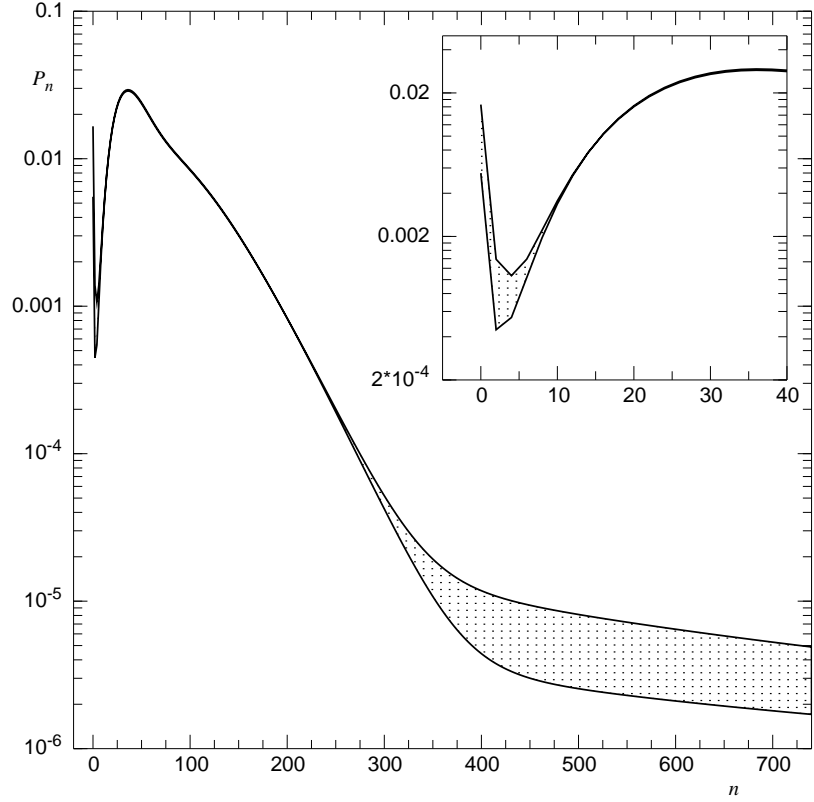}}
  \end{center}
  \caption{$n$ charged particle
	multiplicity distribution $P_n$ expected at 14 TeV in presence
	of a third (maybe hard) component with $\Nbariii=1$, showing one
	shoulder structure and one `elbow' structure. The band illustrates
	the range of values of parameters $\nbariii$, $\kiii$ and $\alphaiii
	= 1 - \alphasoft - \alphasemi$
  discussed in the text.}\label{fig:md3}
  \end{figure}

An ``elbow structure'' is expected in $P_n^{(\text{total})}$  
as the result of the weighted
superposition  of  the NB MD of the semihard component with the gamma 
log-convex MD of the hard one, as shown in Fig.~\ref{fig:md3}.

In the figure are shown in addition to the weighted superposition
of the soft with the  semihard components already seen in 
Ref \cite{combo:prd} the results 
of the weighted superposition  of the semihard and hard components at 14 TeV
c.m.\ energy.

\subsection{The case of few clans}
It should be pointed out that similar results to those discussed so
far for the extreme case $\Nbariii = 1$ can be obtained with a small
average number of clans, provided that $\kiii < 1$, as illustrated in
Table~\ref{tab:1}.
The closer $\kiii$ is to 1, the less evident becomes the elbow structure
discussed above.

\begin{table}
  \caption{Parameters for the third component at 14 TeV, 
		assuming two cases: (a) $\Nbariii=1$ and (b)
		$\kiii=0.5$.}\label{tab:1}
	\begin{center}
  \begin{tabular}{|r|ccccc|}
		\hline
  	& $\alphaiii$	& $\nbariii$ & $\kiii$ & $\Nbariii$ & $\nciii$ \\
		\hline
	  (a) & 0.03	    &   200      &  0.1370 &    1       &   200\\
		    & 0.01      &   700      &  0.1147 &    1       &   700\\
		\hline
		(b) & 0.03      &   200      &   0.5   &   3.62     &   193\\
		    & 0.01      &   700      &   0.5   &   3.00     &    67\\
		\hline
  \end{tabular}
	\end{center}
  \end{table}

\section{Conclusions}

Signals of new physics could be  visible in the tail of 
the $n$ charged particle multiplicity distribution $P_n$
in complete disagreement with Pythia Monte Carlo predictions
under the assumption that the aggregation of the  average number of 
clans  in the semihard component reaches its minimum not asymptotically
but at lower c.m.\ energy. The occurrence of this minimum would suggest
the onset of a third (maybe hard) component to be added to the soft 
and semihard ones discussed in previous work. 

In the paper we examined the main consequences of the occurrence of 
the mentioned situation  already at 14 TeV c.m energy in pp collision,
assuming that the new class of events is between 1 and 3 percent of
the total.

\section*{Acknowledgements}
We would like to thank M. Monteno for providing the Pythia Monte Carlo
simulation results.


\newpage

\end{document}